\documentclass[notitlepage,twocolumn]{revtex4-1}
\usepackage{amssymb}
\usepackage{natbib}
\usepackage{graphicx}
\usepackage{amsmath}
\usepackage[bookmarks = false]{hyperref}
\usepackage{color}
\begin{document}

\title{Hybrid entanglement of three quantum memories with three photons}

\author{Bo Jing$^{1,\,2,\,3,\,*}$}
\author{Xu-Jie Wang$^{1,\,2,\,3,\,*}$}
\author{Yong Yu$^{1,\,2,\,3}$}
\author{Peng-Fei Sun$^{1,\,2,\,3}$}
\author{Yan Jiang$^{1,\,2,\,3}$}
\author{Sheng-Jun Yang$^{1,\,2,\,3}$}
\author{Wen-Hao Jiang$^{1,\,2,\,3}$}
\author{Xi-Yu Luo$^{1,\,2,\,3}$}
\author{Jun Zhang$^{1,\,2,\,3}$}
\author{Xiao Jiang$^{1,\,2,\,3}$}
\author{Xiao-Hui Bao$^{1,\,2,\,3}$}
\author{Jian-Wei Pan$^{1,\,2,\,3}$}

\affiliation{$^1$Hefei National Laboratory for Physical Sciences at Microscale and Department
of Modern Physics, University of Science and Technology of China, Hefei,
Anhui 230026, China}
\affiliation{$^2$CAS Center for Excellence in Quantum Information and Quantum Physics, University of Science and Technology of China, Hefei, Anhui 230026, China}
\affiliation{$^3$CAS-Alibaba Quantum Computing Laboratory, Shanghai 201315, China}
\affiliation{$^*$These two authors contributed equally to this work.}

\begin{abstract}
  Quantum network has significant applications both practically and fundamentally. A hybrid architecture with photons and stationary nodes is highly promising. So far, experimental realizations are limited to two nodes with two photons. Going beyond state of the art by entangling many photons with many quantum nodes is highly appreciated. Here, we report an experiment realizing hybrid entanglement between three photons and three atomic-ensemble quantum memories. We make use of three similar setups, in each of which one pair of photon-memory entanglement with high overall efficiency is created via cavity enhancement. Through three-photon interference, the three quantum memories get entangled with the three photons. Via measuring the photons and applying feedforward, we heraldedly entangle the three memories. Our work demonstrates the largest size of hybrid memory-photon entanglement, which may be employed as a build block to construct larger and complex quantum network.
\end{abstract}

\maketitle

A hybrid architecture with photons and matter-based quantum memories is the most promising approach to build a quantum network~\cite{Cirac1997,Kimble2008,Lvovsky2009}, in which photons act as flying qubits for long-distance transmission, while matter systems act as stationary qubits for storage and manipulation.
A crucial requirement for the matter system is the ability to entangle with photons efficiently. So far, entanglement with a single photon has been demonstrated in many different physical systems. Some systems have also realized the coherent manipulation of two pairs of photon-memory entanglement and generated entanglement between two remote stationary nodes in cold atomic ensembles~\cite{Riedmatten2006,Yuan2008}, single ions~\cite{Moehring2007}, single atom~\cite{Ritter2012,Hofmann2012}, rare-earth ions in crystal~\cite{Usmani2012}, NV centers~\cite{Bernien2013}, and quantum dot~\cite{Delteil2015} etc. Going beyond state of the art, entangling many photons and many quantum memories is highly appreciated, and will enable to build large-scale quantum network with complex topology. For instance, a Greenberger-Horne-Zeilinger (GHZ) type entanglement among three remote quantum memories are the building blocks for two-dimensional quantum repeater~\cite{Wallnofer2016}.

\begin{figure*}[htbp]
\includegraphics{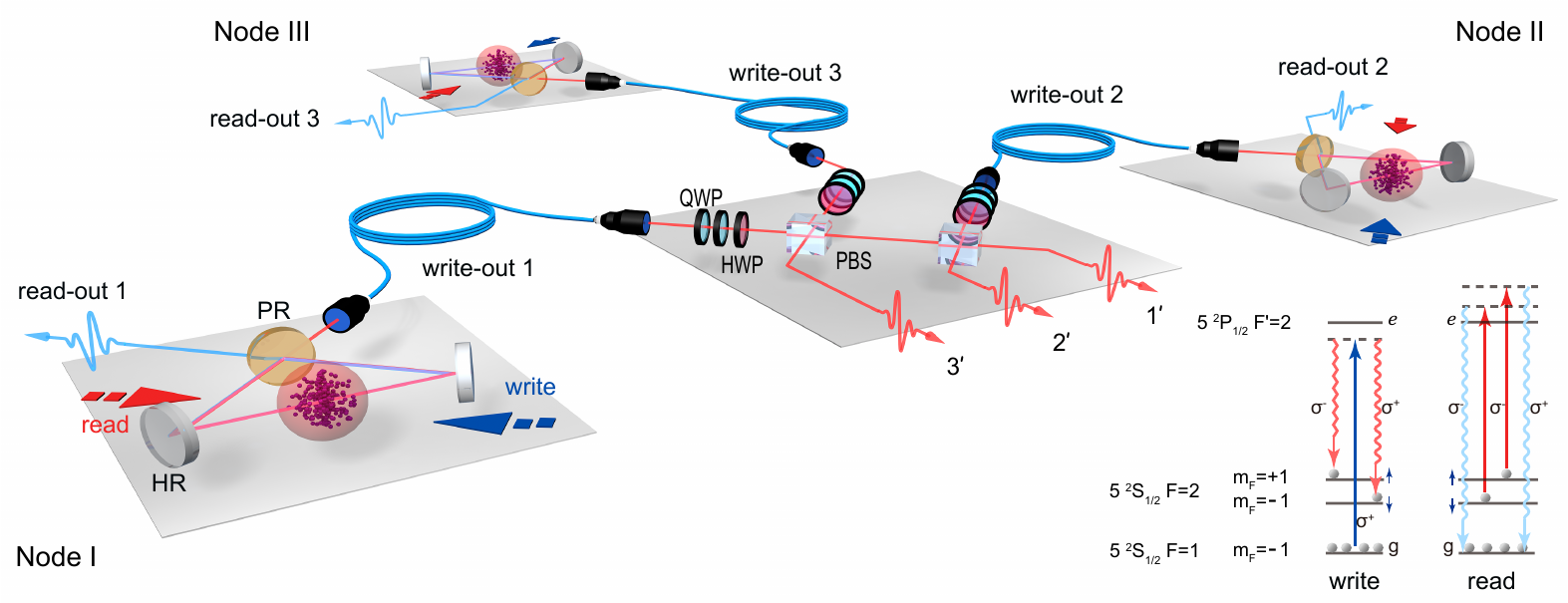}
\caption{\textbf{Experimental layout.} Our experiment involves three quantum nodes. In each node, a setup of atomic-ensemble quantum memory with ring-cavity enhancement is employed. For each setup, Rubidium-87 atoms are initially prepared in a ground state $|g\rangle$, and a pair of atom-photon entanglement is created by making use of dual-rail spontaneous Raman scattering by applying a $\sigma^+$ polarized write beam coupling the transition $|g\rangle\leftrightarrow|e\rangle$. The write-out photons from three setups are sent to a middle station for entanglement connection via quantum interference. Atomic collective states are measured by applying a Raman pulse coupling the transition $\lvert\uparrow\rangle\leftrightarrow\lvert\downarrow\rangle$ for qubit rotation and applying a $\sigma^-$ polarized read beam for optical retrieval and detection. To reduce noise due to leakage and Rayleigh scattered light from the control beams, we configure the write/read beams with detunings of different signs ($\mp$ 40 MHz). HWP: half-wave plate, QWP: quarter-wave plate, PBS: polarizing beam-splitter, PR: partially reflective mirror, HR: highly reflective mirror.}
\label{fig:setup}
\end{figure*}

Laser-cooled atomic ensembles suit very well for the purpose of quantum network~\cite{Sangouard2011}, since single photons can interact with atoms strongly via collective enhancement while good isolation and almost still motion lead to long coherence time~\cite{Yang2015b}. In this paper, we report a six-fold entanglement involving three atomic ensembles and three photons by using a unique cavity-enhanced and filter-free setup of atom-photon entanglement which has a high overall efficiency and a high fidelity. The layout of our experiment is shown in Fig.~\ref{fig:setup}, three similar setups\cite{Bao2012,Yang2015a} are used. For each setup, we make use of spontaneous Raman scattering process\cite{Duan2001} to create a pair of atom-photon entanglement. The entangled state between a write-out photon and the correlated collective atomic state can be written as:
\begin{equation}
|\Psi\rangle=\dfrac{1}{\sqrt{2}} (|\sigma^+\downarrow\rangle+e^{i\varphi (t)}|\sigma^-\uparrow\rangle),
\end{equation}
where $\sigma^{\pm}$ refers to a photonic polarization state, $\lvert\uparrow,\downarrow\rangle$ refers to an atomic collective excitation in a specific internal state, and $\varphi=\varphi_0+2\mu Bt/\hbar $ refers to a phase difference influenced by a bias magnetic field $B$. The atomic states are later retrieved on demand as polarization states of read-out photons by applying a read pulse to verify the atom-photon entanglement.

An important figure of merit for atom-photon entanglement is the overall efficiency $p$, which includes both the entanglement creation efficiency and the verification efficiency. For atomic ensembles, the overall efficiency is defined as $p=p_w \eta_r$, where $p_w$ denotes the probability to detect a write-out photon in each trial and is usually termed as excitation probability, $\eta_r$ denotes the probability to detect a read-out photon conditioned on write-out event and is usually termed as overall retrieval efficiency. In our previous two-node experiment~\cite{Yuan2008}, $p$ is merely around 0.0004, which will lead to a three-node experiment with overall efficiency around $10^{-11}$, which is extremely challenging. To improve $p$, we make use of a ring cavity to enhance the light-matter interaction, which enables us to have a maximal intrinsic retrieval efficiency of $\eta_{int}=0.88$. In previous experiments, it is ubiquitous that external frequency filters (such as etalons and atomic vapor cells) are used to eliminate noise photons, which will inevitably introduce unwanted losses. In this experiment, by setting the detunings of the write and the read beams into different signs, we are able to use the ring cavity itself as a frequency filter. In addition, we use a high order Laguerre-Gaussian (LG) mode for the cavity locking beam, which reduces its leakage into the single-photon channels significantly. As a result, we completely eliminate the use of external frequency filters and get a maximal overall retrieval efficiency of $\eta_r=0.40$. Since the signal-to-noise ratio for the atom-photon entanglement is determined by the ratio of $\eta_r / p_w $, higher $\eta_r$ means that it is possible to operate with a higher $p_w$. By taking above means, we improve the overall efficiency to an average value $p=0.006$, which implies that a three-node efficiency can be improved by about 3 orders of magnitude to $10^{-8}$. Significant further improvement of $p$ is possible by making use of Rydberg blockade~\cite{Saffman2010} to create atom-photon entanglement in a deterministic way~\cite{Li2013g,Li2016}.

\begin{figure}
	\includegraphics[width=\columnwidth]{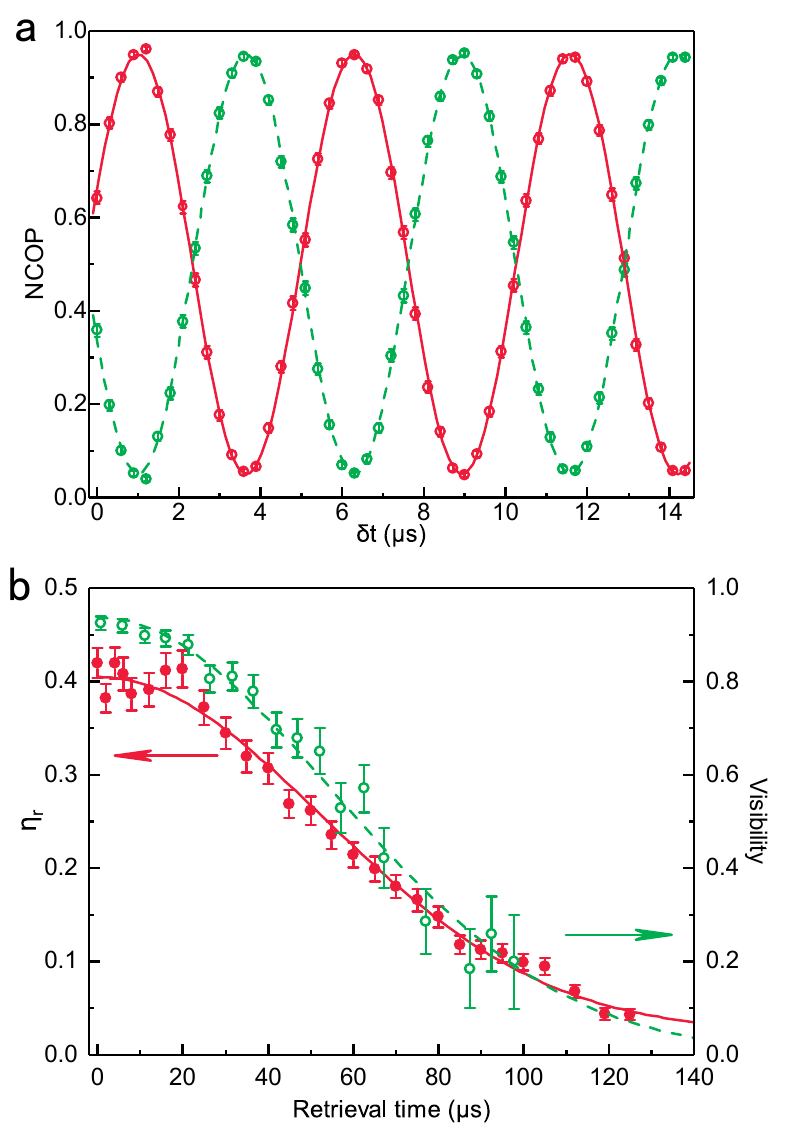}
	\caption{\textbf{Characterization for a single pair of atom-photon entanglement.} \textbf{a.} Normalized coincidence probabilities (NCOP) as a function of Raman pulse delay $\delta t$ relative to the write pulse, with the atomic state measured in the basis of $\lvert\uparrow\rangle\pm\lvert\downarrow\rangle$ and the write-out photon measured in the basis of
$\lvert\sigma^+\rangle\pm\lvert\sigma^-\rangle$. Data points in red (solid line) correspond to parallel correlation, while data points in green (dashed line) correspond to cross correlation. Retrieval time is fixed to 15 $\mu s$. The oscillation has a period of 5.28 $\mu $s which is determined by the Zeeman splitting. \textbf{b.} Measured overall retrieval efficiency $\eta_r$ and entanglement visibility in a superpositional basis as a function of retrieval time. An overall transmission efficiency of 0.65 and a detector efficiency of 0.70 are included in the overall retrieval efficiency.}
	\label{fig2}
\end{figure}

Next we evaluate entanglement fidelity of the atom-photon pair by measuring polarization correlations between the write-out photon and the read-out photon. $\varphi$ can be set as 0 by adjusting the retrieval time. The visibility measured in $|\sigma^\pm\rangle$ basis is 0.901(12), mainly limited by
accidental coincidences. When measuring the visibility in a superpositional basis, a Raman $\pi/2$ pulse is applied to the atomic collective state~\cite{Jiang2016} before retrieval. The results are shown in Fig.~\ref{fig2}a, giving a visibility of 0.901(4), which is nearly the same as the result in $|\sigma^\pm\rangle$ basis. From these measured visibilities, we estimate the fidelity of our atom-photon entanglement as $F=0.926(4)$. We also evaluate coherence of the atomic state by measuring retrieval efficiency and visibility in superpositional basis as a function of retrieval time, with the results shown in Fig.~\ref{fig2}b. The decay of retrieval efficiency gives a 1/e lifetime of 75(2)~$\mu$s. The visibility stays larger than $1/\sqrt{2}$ until 41 $\mu$s. Extension of coherence time to the sub-second regime can be achieved by trapping atoms with a 3-dimensional optical lattice~\cite{Yang2015b}.

To construct a complex quantum network with multiple pairs of atom-photon entanglement, another important requirement is to entangle any two pairs with a high fidelity. In our experiment, entanglement connection is realized through photon interference, and thus it is crucial that the write-out photons from two nodes are highly indistinguishable. Therefore, all the experimental parameters that influence the write-out photon interference, including frequency detuning, pulse shape of the write beams and the bias magnetic field etc., are elaborately optimized during the two-node experiments. Due to the magnetic field, write-out photons in the state of $|\sigma^\pm\rangle$ have a slightly different frequency shift, which will give rise to some indistinguishability and also a detection-time dependent phase shift for the atomic states after entanglement swapping~\cite{Zhao2014}. In our experiment, by flipping the polarization of one write-out photon before interference, different frequency components will not interfere with each other and the detection-time dependent phase shift can be got rid of. This technique enables us to observe an increase of fidelity from 0.81 to 0.85 for a pair of atom-atom entanglement after entanglement swapping in a two-node experiment.

\begin{figure}
	\includegraphics[width=\columnwidth]{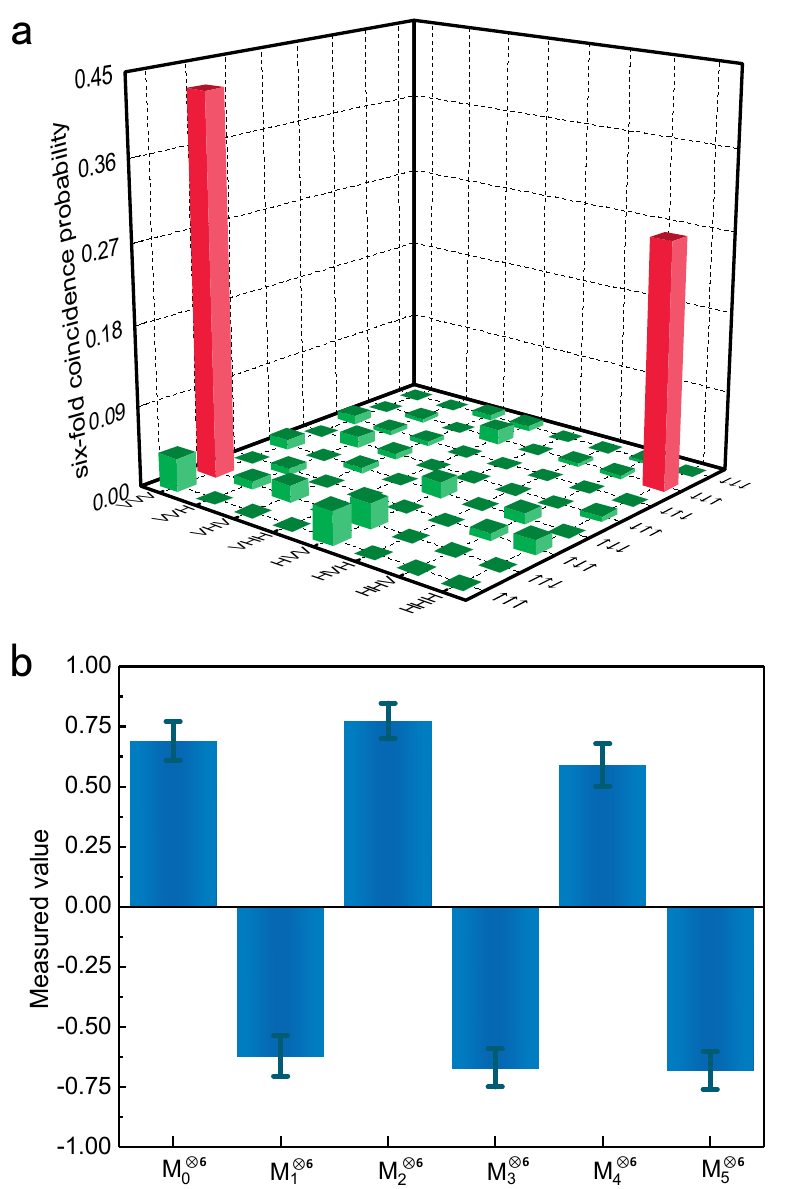}
	\caption{\textbf{Characterization for the six-fold hybrid entanglement $|\text{GHZ}_6\rangle$.} \textbf{a.} Normalized six-fold coincidence probabilities measured in the $|H\rangle/|V\rangle$ basis for the flying write-out photons and in the $\lvert\uparrow\rangle/\lvert\downarrow\rangle$ basis for the atomic states. Result for expected coincidence of $|\text{GHZ}_6\rangle$ is highlighted in red.  \textbf{b.} Measured result for the observables $M_n^{\otimes 6}$. Error bars represent one standard deviation, deduced from propagated Poissonian counting statistics of the detection events. Unbalance in the retrieval efficiency for different atomic states and different setups is calibrated and the six-fold coincidence counting rate is about 6 per hour.}
	\label{6p}
\end{figure}

Based on our efficient atom-photon entanglement and high-fidelity photon interference, we interfere three write-out photons from three setups to generate a six-fold hybrid entanglement of three photons and three quantum memories. The experimental setup is shown in Fig.~\ref{fig:setup}, write-out photons from three separate quantum network nodes are simultaneously sent together for interference, two of them come from node \uppercase\expandafter{\romannumeral1} and \uppercase\expandafter{\romannumeral2} in one lab, and the third one comes from node \uppercase\expandafter{\romannumeral3} located in another lab through an 18 m fiber. Before overlapping at the PBS and interfering, we use waveplates to apply a unitary operation on photon polarizations that obeys the following mapping rules: for photons from node \uppercase\expandafter{\romannumeral1} and node \uppercase\expandafter{\romannumeral2}, $|\sigma^+\rangle\rightarrow |H\rangle$, $|H\rangle\rightarrow|H+V\rangle$, and for photons from node \uppercase\expandafter{\romannumeral3}, $|\sigma^+\rangle\rightarrow |V\rangle$, $|H\rangle\rightarrow|H+V\rangle$. The generated six-fold entangled state can be written as:
\begin{equation*}
|\text{GHZ}_6\rangle=\frac{1}{\sqrt{2}}  (|HHH\downarrow\downarrow\uparrow\rangle+|VVV\uparrow\uparrow\downarrow\rangle),
\label{eqnGHZ6}
\end{equation*}
where the first three qubits represent three photons~(at mode 1$'$, 2$'$, 3$'$) and the last three qubits represent atomic states in node \uppercase\expandafter{\romannumeral1},  \uppercase\expandafter{\romannumeral2},  \uppercase\expandafter{\romannumeral3}. To verify a genuine six-fold hybrid entanglement, we can determine the fidelity of entangled state by decomposing the density operator into observables involving only local measurements as following,
\begin{align*}
\rho_6 =\frac{1}{2}[\,&|HHH\downarrow\downarrow\uparrow\rangle\langle HHH\downarrow\downarrow\uparrow| \\ +&|VVV\uparrow\uparrow\downarrow\rangle\langle VVV\uparrow\uparrow\downarrow|\,] +\frac{1}{12}\sum_{n=0}^{5} (-1)^nM_n^{\otimes 6},
\end{align*}
where $M_n=\cos (n\pi/6)\sigma_x+\sin (n\pi/6)\sigma_y$ are measured in different bases and $\sigma_{x,y,z}$ are Pauli matrices after defining the eigen bases $|H\rangle/|V\rangle $ for photons and $\lvert\uparrow\rangle/\lvert\downarrow\rangle$ for atoms. After converting the atomic states to the polarization states of read-out photons, we first measure the six-fold coincidence in the eigen bases. The six-fold probability is 0.43 for state $|VVV\uparrow\uparrow\downarrow\rangle$, 0.28 for $|HHH\downarrow\downarrow\uparrow\rangle$  and 0.29 for other 62 state combinations, as shown in Fig.~\ref{6p}a. Afterwards, we measure the other six observables $M_n^{\otimes 6}$ as shown in Fig.~\ref{6p}b. With these results, we can estimate a six-fold entanglement with a fidelity of $0.686\pm0.026$, exceeding the threshold 0.5 by 7 standard deviations, and proving the existence of a genuine six-fold hybrid entanglement, which is the largest size of atom-photon entanglement reported so far.

\begin{figure}
	\includegraphics[width=\columnwidth]{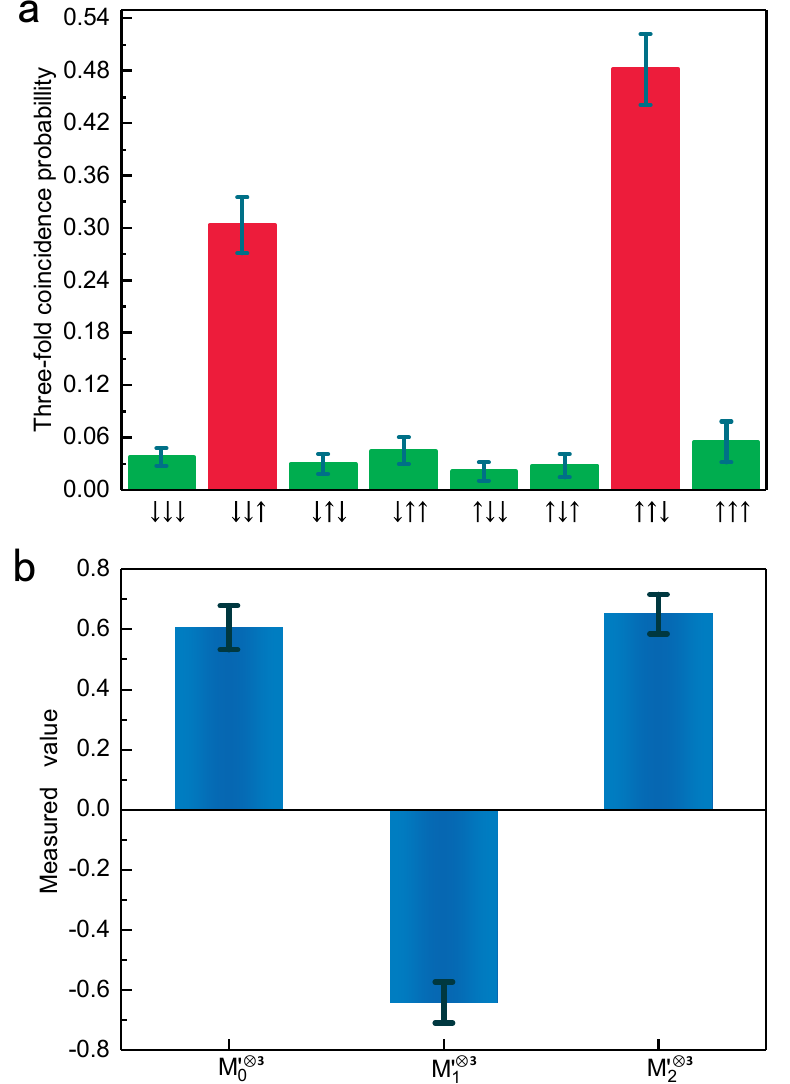}
    \caption{\textbf{Characterization for the three-memory entanglement $|\text{GHZ}^+_3\rangle$.} \textbf{a.} Normalized three-fold coincidence probabilities measured in the $\lvert\uparrow\rangle/\lvert\downarrow\rangle$ basis for the atomic states. Result for expected coincidence of $|\text{GHZ}^+_3\rangle$ is highlighted in red.  \textbf{b.} Measured result for the observables $M_n^{'\otimes 3}$. Error bars represent one standard deviation, deduced from the propagated Poissonian counting statistics of the detection events. Unbalance in the retrieval efficiency for different atomic states and different setups is calibrated.}
\label{3en}
\end{figure}

Furthermore, by making projective measurements on photons 1$'$, 2$'$, 3$'$ at $|D\rangle/|A\rangle$ basis ($|D\rangle/|A\rangle$ = $\frac{1}{\sqrt{2}}|H\pm V\rangle$), we experimentally demonstrate entanglement of three quantum memories. Coincident detection of $|DDD\rangle$, $|AAD\rangle$, $|DAA\rangle$, $|ADA\rangle$  will project the atomic entangled state to $|\text{GHZ}_3^+\rangle=\frac{1}{\sqrt{2}} (\lvert\downarrow\downarrow\uparrow\rangle+|\!\uparrow\uparrow\downarrow\rangle)$. While detection of  $|ADD\rangle$, $|DAD\rangle$, $|DDA\rangle$, $|AAA\rangle$ will project entangled state to $|\text{GHZ}_3^-\rangle=\frac{1}{\sqrt{2}} (\lvert\downarrow\downarrow\uparrow\rangle-\lvert\uparrow\uparrow\downarrow\rangle)$, then we apply a feed-forward operation to restore the atomic state back to $|\text{GHZ}_3^+\rangle$. Afterwards, atomic states are converted to polarization states of read-out photons and detected. To measure the state fidelity via local measurements, we decompose the corresponding density matrix as following:
\begin{equation*}
\rho_3=\frac{1}{2}\left (|\!\downarrow\downarrow\uparrow\rangle\langle \downarrow\downarrow\uparrow\!|\!+|\!\uparrow\uparrow\downarrow\rangle\langle \uparrow\uparrow\downarrow\!| \right) +\frac{1}{6}\sum_{n=0}^{2} (-1)^nM_n^{'\otimes 3}.
\end{equation*}
Here $M_n^{'}\text{ stands for } \cos (n\pi/3)\sigma_x+\sin (n\pi/3)\sigma_y$. The result (Fig.~\ref{3en}a) tells that the coincidence probability is 0.30 for state $\lvert\downarrow\downarrow\uparrow\rangle$, 0.48 for $\lvert\uparrow\uparrow\downarrow\rangle$ and a total probability of 0.22 for other six combinations. Then $M_n^{'\otimes 3}$ are measured to identify the state is indeed entangled (Fig.~\ref{3en}b). With these results, we calculate the three-fold entanglement fidelity as 0.709$\pm$0.026. Thus, with a high statistical significance (exceeding the bound of 0.5 by 8 standard deviations), we create entanglement among three atomic-ensemble quantum memories. We note that the photonic projective measurement may be in principle used as heralding signal which tells the success of three-fold atomic entanglement. While in our experiment, conditioned on this heralding signal, the chance that three atomic ensemble are in the expected state is around 0.14, which can be improved to nearly unity if Rydberg-based nonlinearity is employed in each ensemble to inhibit high-order excitations~\cite{Li2013g,Li2016}. We also note that our experiment is very robust against losses and phase fluctuations, and can be extended to long-distance separated quantum nodes, which is prohibited for previous entangling experiments, such as sharing a single excitation among multiple ensembles~\cite{Choi2010,Pu2018} or storing multipartite continuous-variables entanglement with atomic ensembles~\cite{Yan2017}.

In conclusion, based on our efficient source of atom-photon entanglement with cavity enhancement,  we report realization of the largest hybrid entanglement involving three quantum memories and three photons. By detecting the photons, three quantum memories are entangled heraldedly, which may be used as building blocks to construct a 2D quantum repeater~\cite{Wallnofer2016}, and build quantum networks with complex topology~\cite{Barrett2010}. Extension of memory lifetime to the second regime is straight forward by using optical lattice~\cite{Yang2015b}, which will enable a deterministic creation of entanglement within lifetime. By harnessing Rydberg blockade~\cite{Saffman2010}, the creation of atom-photon entanglement may become deterministic, which will enable a much higher rate for the creation of multi-node entanglement, and also eliminating the unwanted events during heralding. Extension of distances for memory separations can be realized by converting photon wavelength to telecom-band and using low-loss fiber for transmission~\cite{Maring2017}. Thus our experiment paves a way towards building a long-distance quantum network, and will enable a number of interesting applications over it, such as multi-party quantum communication, distributed quantum computing~\cite{Jiang2007f}, synchronization of atomic clocks~\cite{Komar2014}, and construction of long-baseline telescopes~\cite{Gottesman2012}.

This work was supported by National Key R\&D Program of China (No. 2017YFA0303902), Anhui Initiative in Quantum Information Technologies, National Natural Science Foundation of China, and the Chinese Academy of Sciences.

\bibliography{myref}

\clearpage

\onecolumngrid

\appendix
\section{Time sequence}
Our experiment runs with a period of 21 ms. In each cycle, the starting 18 ms are used for MOT loading, molasses cooling and optical pumping. Afterwards the memory phase starts with a duration of 3 ms, during which we repeat the write trials. Each write trial requires 4.7 $\mu$s, and maximally 622 trials are allowed for one atomic loading. Once three write-out photons are detected, we apply the Raman pulse and retrieve the atomic states for entanglement verification.

\section{Performing measurement in a superpositional basis.}
When measuring the atomic state in a superpositional basis, we do not directly retrieve the atomic state to a read-out photon's polarization, since the retrieval efficiency is slightly state-dependent. Instead, we apply a Raman pulse which rotates the atomic qubit before retrieval. The read-out photon is always measured in $|\sigma^\pm\rangle$ basis. Thanks to the different phase evolution speed for the atomic states, by adjusting time delay and width for the Raman pulse, we can measure the atomic state in an arbitrary basis of $\alpha\lvert\downarrow\rangle \pm e^{i\varphi '}\beta\lvert\uparrow\rangle$.

\section{Detailed experimental setup for a single quantum memory}
\begin{figure}[htbp]
  	\includegraphics[width=0.8\textwidth]{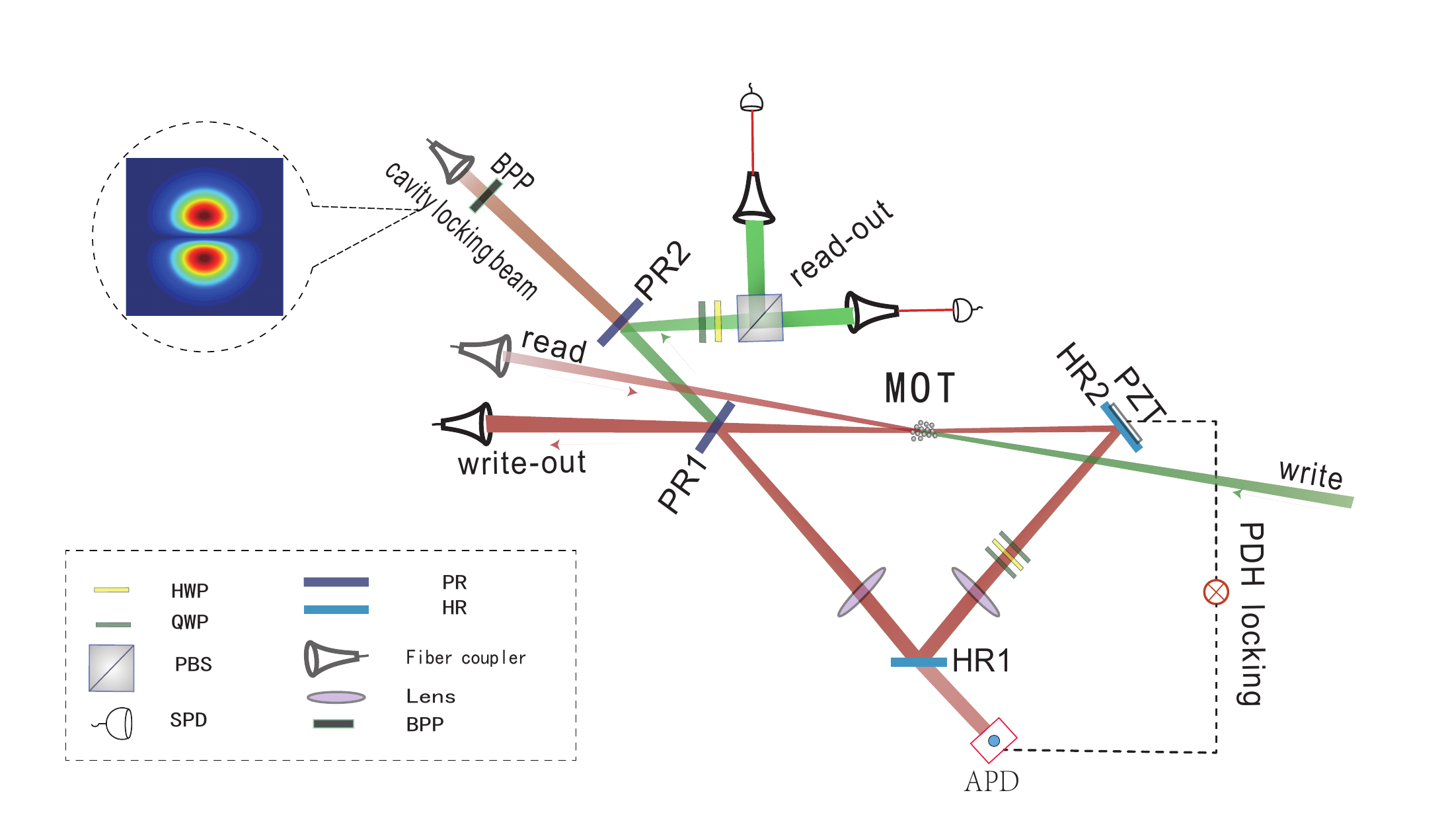}
  	\caption{\textbf{Experimental setup for a single quantum memory}. HWP: half wave plate, QWP: quarter wave plate, PBS: polarizing beam splitter, SPD: single photon detector, PR: partial reflective mirror, HR: high reflective mirror, BPP: binary phase plate.}
  	\label{energylevel}
\end{figure}

As shown in Fig.~\ref{energylevel}, the counterpropagating write and read pulse intersect with the cavity mode through a cold atomic ensemble with an angle of $2.5^\circ$. For the write and read modes,  the beam waist radius is 300 $\mu$m, while for the write-out and read-out modes, the beam waist radius is 90 $\mu$m. The ring cavity mainly consists of two lenses ($f$ = 250 mm), two highly reflecting mirrors (HR1 and HR2) and one partially reflecting mirror PR1 (R $\sim$ 80\%) as signal photon coupling ports. Two quarter wave plates and one half wave plate are used for polarization compensation after each round trip in the cavity. To ensure the cavity length stable, we use the Pound-Drever-Hall (PDH) locking method to lock the cavity. The cavity-locking beam with $H$ polarization is 630 MHz blue detuned from the upper energy level $|e\rangle$  combined with the read-out channel through a PR2 (R $\sim$ 95\%), then the transmission signal from a HR1 (R $>$ 99.9\%) is detected and sent to the PDH locking module, and feedback drive the mirror HR2 with a piezoelectric transducer (PZT) for cavity stabilization. The measured free spectral range (FSR) of the cavity is about 466 MHz and finesse $\mathcal{F} = 23.5$.

\section{Contribution of accidental coincidences for one-pair entanglement visibility}
After applying the read pulse to retrieve the atomic state, the entangled state between a write-out and a read-out photon can be described as :
 \begin{equation}
|\Psi\rangle=\dfrac{1}{\sqrt{2}}(|RL\rangle+|LR\rangle).
\end{equation}
For $N$ experimental trials, measured coincidence counts between the write-out and read-out photons after calibration are $n_{ij}$, where $i,j=\{R,L\}$ stands for polarization of the two photons. The raw visibility in the $|R\rangle/|L\rangle$ basis is calculated as:
\begin{equation}
V_{RL}=\dfrac{n_{RL}+n_{LR}-n_{LL}-n_{RR}}{n_{RL}+n_{LR}+n_{LL}+n_{RR}}.
\end{equation}
To calculate accidental coincidence counts, we need to know single counts for the write-out photon ($n_{woR}$ and $n_{woL}$) and single counts for the read-out photon ($n_{roR}$ and $n_{roL}$).
We carry out the subtraction of accidental coincidences as following:
\begin{equation}
\begin{aligned}
n'_{RL}&=n_{RL}-n_{woR}n_{roL}/N\\
n'_{LR}&=n_{LR}-n_{woL}n_{roR}/N\\
n'_{LL}&=n_{LL}-n_{woL}n_{roL}/N\\
n'_{RR}&=n_{RR}-n_{woR}n_{roR}/N.
\end{aligned}
\end{equation}
The new visibility is described as:
\begin{equation}
V'_{RL}=\dfrac{n'_{RL}+n'_{LR}-n'_{LL}-n'_{RR}}{n'_{RL}+n'_{LR}+n'_{LL}+n'_{RR}}.
\end{equation}
By applying this procedure, the visibility in $|R\rangle/|L\rangle$ basis gets increased from 0.90 to 0.98. By applying a similar procedure for the result measured in the superpositional basis, the visibility gets increased from 0.90 to 0.97. This analysis implies that accidental coincidence contributes dominantly to the infidelity of our atom-photon entanglement.

\begin{figure}[t]
	\includegraphics[width=0.8\textwidth]{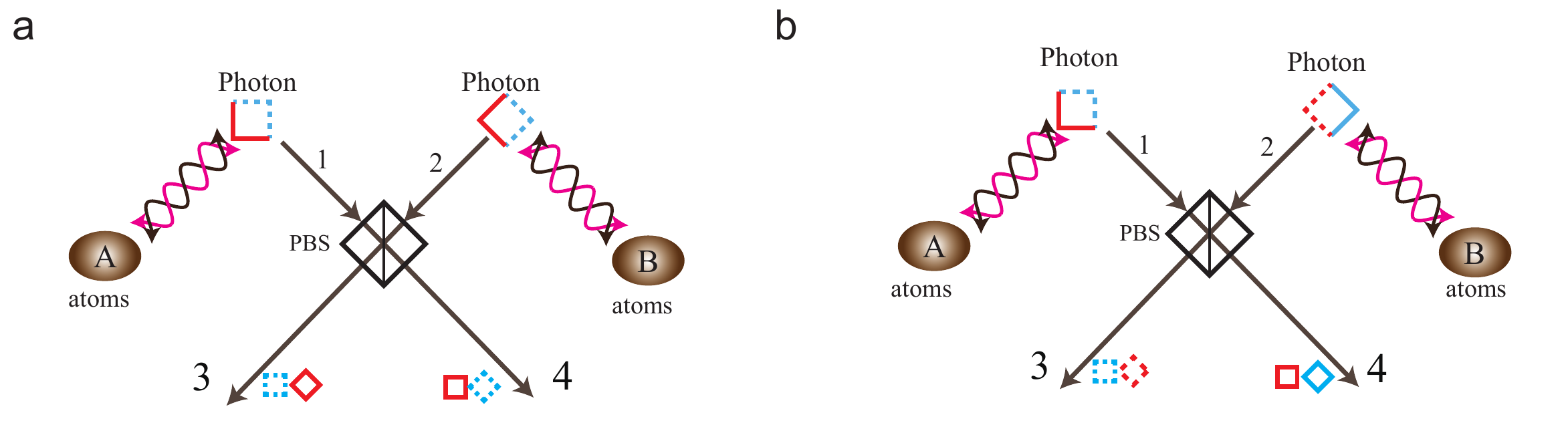}
	\caption{\textbf{Schematic for generating atom-atom entanglement}. \textbf{a.} Photon at 1  and photon at 2 have same polarizations for same frequency. \textbf{b.} Photon at 1  and photon at 2 have different polarizations for same frequency by flipping polarization of photon at 2. Different frequency is presented in solid ($\omega_+$) and dot ($\omega_-$). Different polarizations are presented in red (H) and blue (V).}
	\label{2node}
\end{figure}

\section{Benefit by flipping polarization before interference with two pairs of entanglement}
It is crucial for quantum interference that photons need to be highly indistinguishable. In our experiment however, duo to Zeeman splitting, the write-out photon's frequency is dependent on its polarization ($\omega_+$ for $\sigma^+$ polarization and $\omega_-$ for $\sigma^-$ polarization). By applying a unitary operation on photon's polarization ($|\sigma^+\rangle\rightarrow |H\rangle$, $|H\rangle\rightarrow|H+V\rangle$), the combined photonic state can be denoted as $|H,\omega_+\rangle$ and $|V,\omega_-\rangle$. Therefore, when performing a two-pair experiment (Fig.~\ref{2node}a), the atom-photon entanglement before interference can be rewritten as:
\begin{equation}
\begin{aligned}
|\Psi\rangle_{1A}&=\dfrac{1}{\sqrt{2}}(|H,\omega_+\rangle_1\lvert\downarrow\rangle_A+|V,\omega_-\rangle_1\lvert\uparrow\rangle_A)\\
|\Psi\rangle_{2B}&=\dfrac{1}{\sqrt{2}}(|H,\omega_+\rangle_2\lvert\downarrow\rangle_B+|V,\omega_-\rangle_2\lvert\uparrow\rangle_B).
\end{aligned}
\end{equation}
To entangle atomic state A and B, photons at 1, 2  are sent together for a joint Bell state measurement (BSM), the case that two photons go to different output ports 3 and 4 can be post-selected (assuming detection time  are $t_3$, $t_4$), giving a joint state after PBS as:
\begin{equation}\label{noflip}
|\Psi\rangle_a=\dfrac{1}{\sqrt{2}}(|H,\omega_+\rangle_4|H,\omega_+\rangle_3\lvert\downarrow\rangle_A\lvert\downarrow\rangle_B+|V,\omega_-\rangle_3|V,\omega_-\rangle_4\lvert\uparrow\rangle_A\lvert\uparrow\rangle_B)
\end{equation}
To study the influence of partial distinguishability, we decompose the frequency components in the temporal domain as following:
\begin{equation}\label{fouri}
\begin{aligned}
	|\omega_+\rangle&=\int dte^{-i\omega_+t}f(t)|t\rangle\\
	|\omega_-\rangle&=\int dte^{-i\omega_-t}g(t)|t\rangle,
\end{aligned}
\end{equation}
where $f(t)$, $g(t)$ denote the temporal shape for the $\omega_+$ photon and $\omega_-$ photon respectively. Substituting
 Eq.~\ref{fouri} to Eq.~\ref{noflip}, we get:
\begin{equation}
|\Psi\rangle_a=\dfrac{1}{\sqrt{2}}\int\int dt_3dt_4e^{-i\omega_+(t_3+t_4)}f(t_3)f(t_4)[|\Psi\rangle_{AB}^+\otimes|\Phi\rangle^++|\Psi\rangle_{AB}^-\otimes|\Phi\rangle^-],
\end{equation}
where the photonic part $|\Phi\rangle^{\pm}=\dfrac{1}{\sqrt{2}}(|H,t_4\rangle_4|H,t_3\rangle_3\pm|V,t_3\rangle_3|V,t_4\rangle_4)$ can be distinguished by making further projective measurements on photons at 3, 4 in $|D\rangle/|A\rangle$ basis, and the atomic part is accordingly projected to
\begin{equation}\label{atom}
  |\Psi\rangle_{AB}^\pm=\frac{1}{\sqrt{2}}(\lvert\downarrow\rangle_A\lvert\downarrow\rangle_B\pm e^{-i(\omega_--\omega_+)(t_3+t_4)}\dfrac{g(t_3)g(t_4)}{f(t_3)f(t_4)}\lvert\uparrow\rangle_A\lvert\uparrow\rangle_B).
\end{equation}
If the two frequency components ($|\omega_+\rangle$ and $|\omega_-\rangle$) are totally indistinguishable, namely $\omega_+=\omega_-$ and $f(t)=g(t)$, the atomic state $|\Psi\rangle_{AB}^\pm$ will be a maximally entangled state. In our experiment however, the photon detection time varies from event to event. On average, it will result in a partially mixed state for the final atomic entanglement. Additional distinguishability in the pulse shape ($f(t) \neq g(t)$) will further reduce the purity and fidelity for the final atomic entanglement. To solve this problem, we need to eliminate the interference between different frequency components. In our experiment, it is simply done by flip the polarization for one write-out photon before interference (Fig.~\ref{2node}b). By doing this, the entangled state $|\Psi\rangle_{2B}$ now becomes:
\begin{equation}
|\Psi\rangle_{2B}=\dfrac{1}{\sqrt{2}}(|V,\omega_+\rangle_2\lvert\downarrow\rangle_B+|H,\omega_-\rangle_2\lvert\uparrow\rangle_B
\end{equation}
Similarly, we post-select the case that two photons go to different output ports 3 and 4, which gives a joint state of:
\begin{equation}\label{flip}
\begin{aligned}
|\Psi\rangle_b=&\dfrac{1}{\sqrt{2}}(|H,\omega_+\rangle_4|H,\omega_-\rangle_3\lvert\downarrow\rangle_A\lvert\uparrow\rangle_B+|V,\omega_-\rangle_3|V,\omega_+\rangle_4\lvert\uparrow\rangle_A\lvert\downarrow\rangle_B)\\
=&\dfrac{1}{\sqrt{2}}|\omega_+\rangle_4|\omega_-\rangle_3(|H\rangle_4|H\rangle_3\lvert\downarrow\rangle_A\lvert\uparrow\rangle_B+|V\rangle_3|V\rangle_4\lvert\uparrow\rangle_A\lvert\downarrow\rangle_B)\\
=&\dfrac{1}{2\sqrt{2}}|\omega_+\rangle_4|\omega_-\rangle_3[(\lvert\downarrow\rangle_A\lvert\uparrow\rangle_B+\lvert\uparrow\rangle_A\lvert\downarrow\rangle_B)\otimes(|H\rangle_4|H\rangle_3+|V\rangle_3|V\rangle_4)\\
&+(\lvert\downarrow\rangle_A\lvert\uparrow\rangle_B-\lvert\uparrow\rangle_A\lvert\downarrow\rangle_B)\otimes(|H\rangle_4|H\rangle_3-|V\rangle_3|V\rangle_4)],
\end{aligned}
\end{equation}
where the frequency components do not interfere each other anymore and the internal phase for the joint atomic state is not dependent on the detection time. By applying this operation in our experiment, the atomic entanglement's fidelity in a two-node experiment is improved from 0.81 to 0.85.

\end{document}